# Spectral-control-assisted Kerr-lens mode-locking in a $Tm:Y_2O_3$ ceramic laser

ZEYU GONG,[1,2] JING WANG,[1,2] PANQIANG KANG,[1] WEICHAO YAO,[1,*] YUJIE PENG,[1] JUN WANG,[3,5] DINGYUAN TANG,[4] YUXIN LENG[1,2,6]

[1]State Key Laboratory of Ultra-intense Laser Science and Technology, Shanghai Institute of Optics and Fine Mechanics, Chinese Academy of Sciences, Shanghai 201800, China

[2]Center of Materials Science and Optoelectronics Engineering, University of Chinese Academy of Sciences, Beijing 100049, China

[3]Jiangsu Key Laboratory of Advanced Laser Materials and Devices, Jiangsu Normal University, Xuzhou 221116, China

[4]Future Technology School, Shenzhen Technology University, Shenzhen 518118, China

[5]junwang@jsnu.edu.cn

[6]lengyuxin@mail.siom.ac.cn

*wyao@siom.ac.cn

**We report a soft-aperture Kerr-lens mode-locked $Tm:Y_2O_3$ ceramic laser stabilized by suppressing wavelength competition under high population-inversion conditions. The laser delivers an average output power of 0.5 W with a pulse duration of 112 fs at a repetition rate of 90 MHz and a central wavelength of 2.08 μm. Further optimization for shorter pulses yields a pulse duration of 107 fs, through with a reduced output power of 0.22 W. These results demonstrate that spectral-control is an effective approach for achieving stable mode-locking in such laser with strong spectral competition.**

Ultrafast solid-state lasers operating at 2 μm have broad applications in gas sensing, materials processing, and high-harmonic generation. In this regard, mode-locked Tm solid-state lasers provides an attractive approach, since $Tm^{3+}$-doped materials typically exhibit ultra-broad gain bandwidth that enable femtosecond pulses generation [1]. Femtosecond pulses generation have been realized based on various disordered materials, including $Tm:GdScO_3$ [2], Tm:CNNGG [3], Tm:CALGO [4], and $Tm^{3+}$-doped sesquioxide [5-7], etc. As a prominent family of gain media for femtosecond pulse generation, $Tm^{3+}$-doped sesquioxide offers excellent thermal properties, making them highly promising candidates for the development of high-power ultrafast laser system. To date, SESAM mode-locking has been the most widely used technical approach for Tm-based ultrafast lasers [5-7].

Although SESAM mode-locking offers simple implementation and supports high average power generation, its intrinsically slow response time and limited bandwidth can restrict the generation of ultrashort pulses. In contrast, Kerr-lens mode-locking (KLM) is well suited to the formation of a shorter pulse duration, owing to its large modulation depth, ultrafast dynamic response mechanism, and unrestricted spectral bandwidth. However, only a limited number of KLM demonstrations have been reported for Tm sesquioxide lasers [8, 10–12]. A commonly challenge is that a smooth gain spectrum can only be obtained under low population-inversion conditions. High population inversion will reshape the gain spectrum into a multi-peak profile and induce a blue shift of the gain peak. Such spectral shift overlaps with the water vapor absorption band, which severely compromises the stability of femtosecond pulse generation [9, 13]. The problem is particularly pronounced in KLM laser system because the cavity is typically operated near the edge of the stability range, allowing the gain medium to reach a high population inversion before mode-locking is initiated. A substantial wavelength shift not only hinders mode-locking initiation but also limits the use of high-transmission output couplers for power scaling.

To control the laser wavelength during KLM operation in Tm sesquioxide lasers, we designed a dichroic mirror with high reflectivity at the laser wavelength under low population inversion, and high transmission at the laser wavelength under high population inversion. This design enables precise control of the laser spectrum and constrains the lasing wavelength to above 2000 nm. Using this approach, we achieved KLM in a $Tm:Y_2O_3$ ceramic laser with a pulse duration of 112 fs, an average power of 0.5 W, at a repetition rate of 90 MHz and a central wavelength of 2.08 μm. This spectral-control technique provides a universal and reliable solution for stabilizing KLM operation in Tm laser systems.

Figure 1 shows the experimental setup of the mode-locked $Tm:Y_2O_3$ ceramic laser. A 2 at.% $Tm:Y_2O_3$ ceramic with a length of 3 mm was positioned at Brewster's angle. The ceramic has an aperture of 3 mm ×3 mm, and both end surfaces of the crystal were finely polished without anti-reflection coatings. The ceramic was pumped by a linearly polarized 1635 nm Raman fiber laser with a maximum output power of 20 W. The pump beam radii in the sagittal and tangential plane in the ceramic were 32 μm and 60 μm, respectively. The small-signal absorption of the $Tm:Y_2O_3$ ceramic at low pump power (<300 mW) was ~80% under non-lasing condition. Two concave mirrors (M1 and M2) with high transmission at 1600-1700 nm and high reflectivity at 1950-2200 nm were used, both have radius of curvature (RoC) of -100 mm. The high-reflectivity (HR) mirrors and chirped mirrors (CM1-CM3) in the cavity also exhibited high reflectivity at 1900-2200 nm. Different output couplers with transmittances ($T_{OC}$) of 1%, 2%, 3%, 5%, and 7% at 2000-2200 nm were used to optimize the output power. A 45° dichroic mirror (DM) was used to control the laser wavelength. The total round-trip group-delay dispersion of the cavity was around -3000 $fs^2$ to support soliton mode-locking. The total cavity length was approximately 1.66 m, corresponding to a repetition rate of 90 MHz.

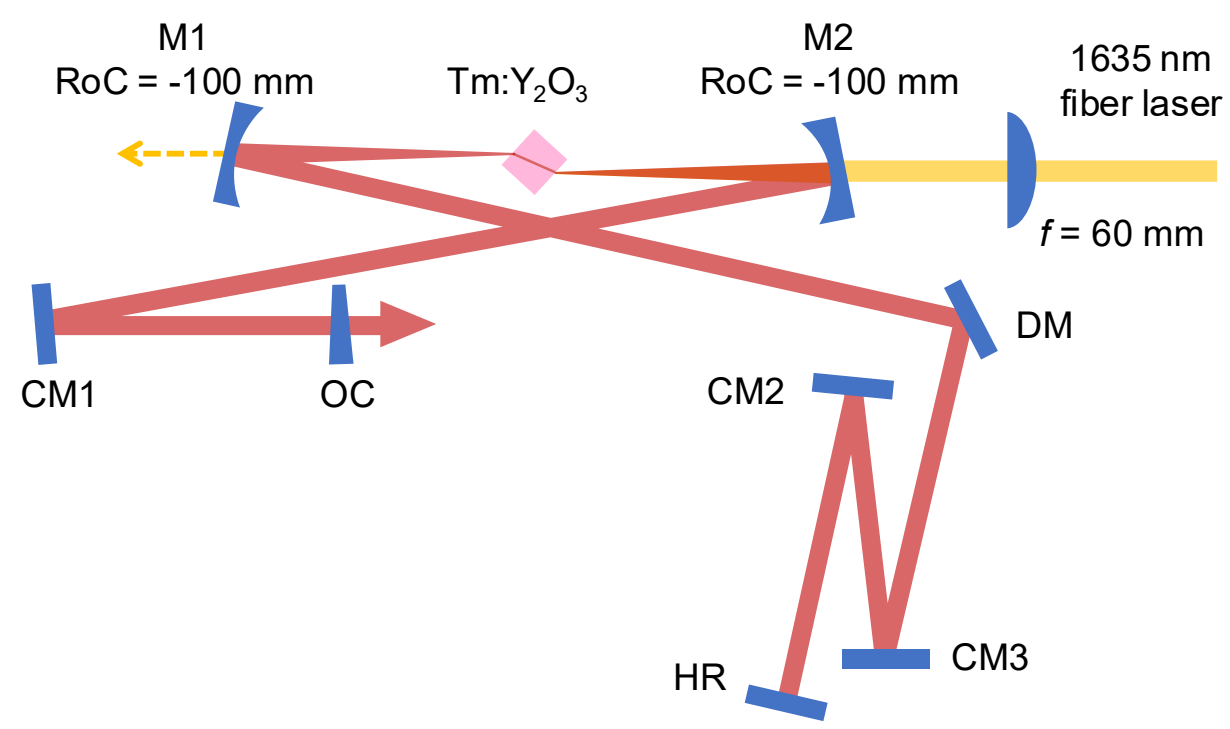

Fig. 1. Experimental setup of the KLM $Tm:Y_2O_3$ ceramic laser. M1 and M2: concave mirrors. RoC: radius of curvature. CM1-CM3: chirped mirrors, -500 $fs^2$ per bounce. OC: output coupler. DM: dichroic mirror. HR: high-reflectivity mirror.

Figure 2 shows the reflectivity spectra of the two dichroic mirrors, DM1 and DM2, together with the gain cross-section of the $Tm:Y_2O_3$ ceramic. DM1 exhibits high reflectivity at 1900-2200 nm, whereas DM2 features high transmission at 1900-1950 nm and high reflectivity at 2020-2200 nm. The gain cross-section shows two wavelength peaks at 1950 nm and 2050 nm under a high population inversion ratio (β). This spectral structure explains the shift of the $Tm:Y_2O_3$ laser wavelength from 2050 nm to 1950 nm when no wavelength-selective element is used, i.e., when the broadband reflective mirror DM1 is employed.

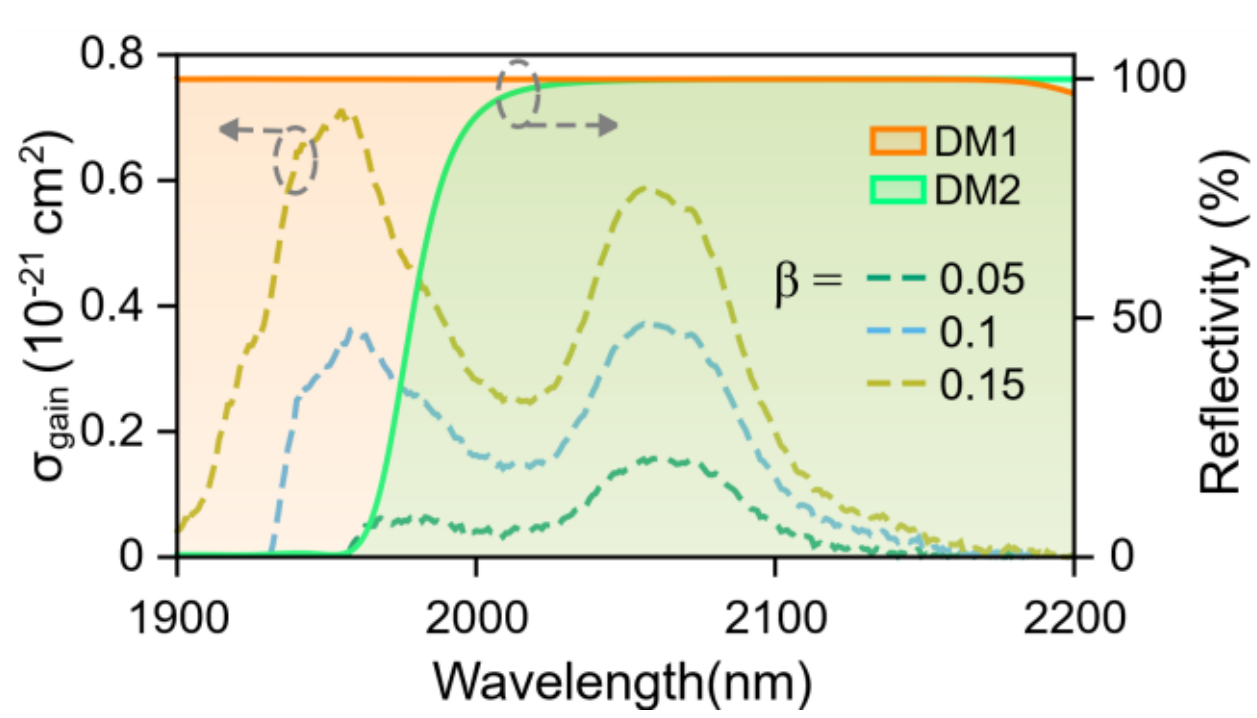

Fig. 2. Reflectivity spectra of DM1 and DM2, and the gain cross-section of the $Tm:Y_2O_3$ ceramic. β: population inversion ratio.

First, we optimized the laser cavity for maximum continuous-wave (CW) output power with DM1. Figure 3(a) shows the output power as a function of the absorbed pump power for different $T_{OC}$s. The slope efficiencies were 12.5%, 19.7%, 23.8%, 28.4%, and 28.8% for $T_{OC}$ of 1%, 2%, 3%, 5%, and 7%, respectively. At an absorbed pump power of 16 W, a maximum output power of 4.8 W was achieved with 7% $T_{OC}$, corresponding to an optical-to-optical efficiency of 30%. No power saturation was observed, suggesting that the output power could be further scaled with a higher available pump power. Laser spectra were recorded using a spectrometer (WaveScan, A.P.E. GmbH). Figure 3(b) shows the laser spectra with different $T_{OC}$s. The central wavelength remained stable at ~2050 nm for $T_{OC}$s of 1%, 2%, 3%, and 5%, but shifted to ~1950 nm with 7% $T_{OC}$, indicating that the population inversion ratio exceeded 10% at this output coupling.

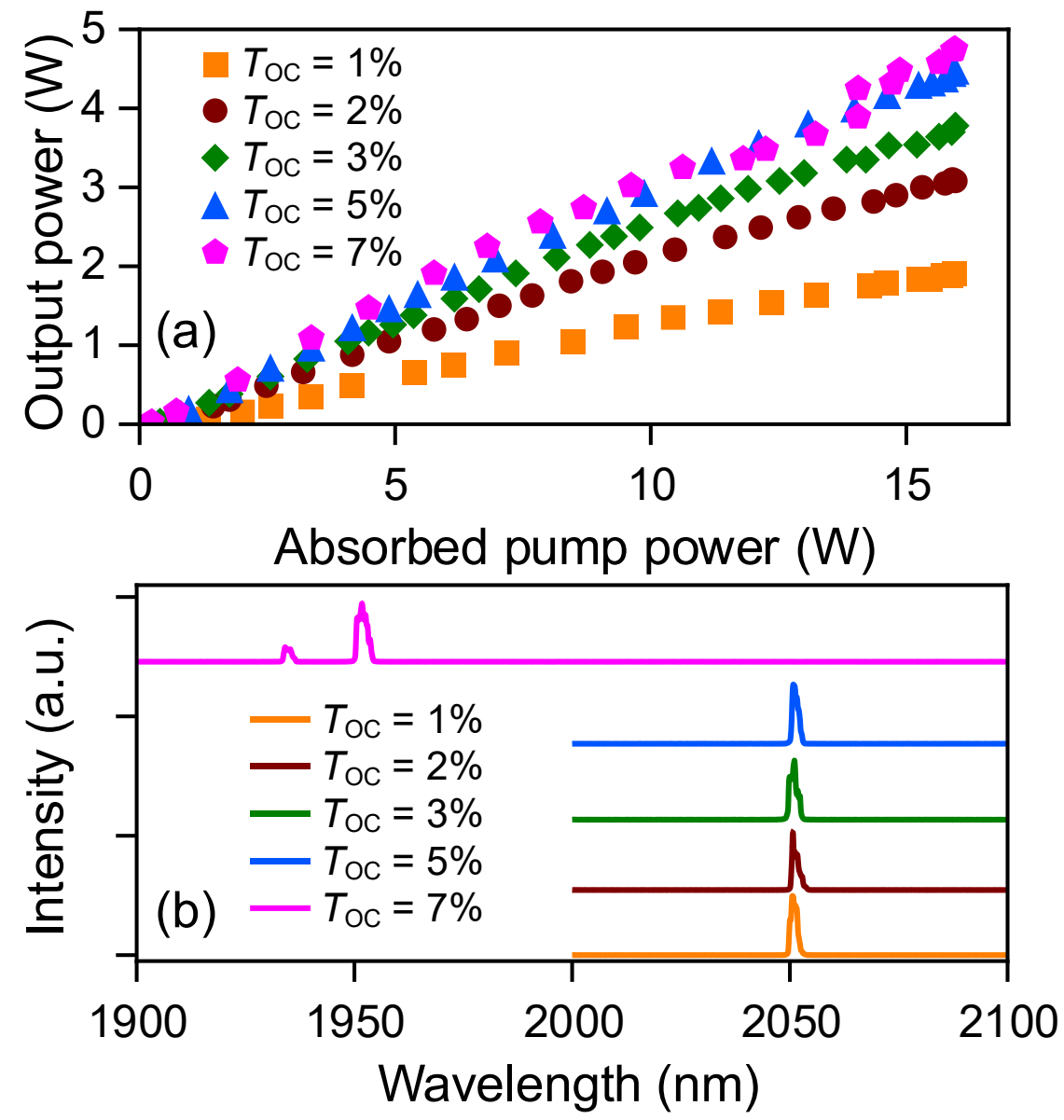

Fig. 3. CW laser output results. (a) Output power as a function of the absorbed pump power for different $T_{OC}$s. (b) Laser spectra for different $T_{OC}$s.

To initiate mode-locking, $T_{OC}$ was set to 1% because mode-locking is more readily achievable at low output coupling. The cavity mirror M1 was moved toward the crystal until the output power decreased rapidly. As a result, the single-wavelength laser emission at 2050 nm switched to dual-wavelength operation at 2050 nm and 1950 nm [Fig. 4(a)], or even single-wavelength emission at 1950 nm. In this case, stable mode-locking could not be achieved despite careful cavity alignment and the application of external mechanical perturbations. In the best case, only incomplete mode-locking was obtained, with the laser spectrum dominated by a strong CW component [Fig. 4(b)]. This behavior is primarily attributed to insufficient gain competition between the CW and the mode-locked pulses, which prevents the complete suppression of the CW component. In general, moving the cavity working point closer to the edge of the stability range can effectively strengthen gain competition. However, this will also increase the population inversion and enhances the blue shift of the lasing wavelength, which is detrimental to stable mode-locking.

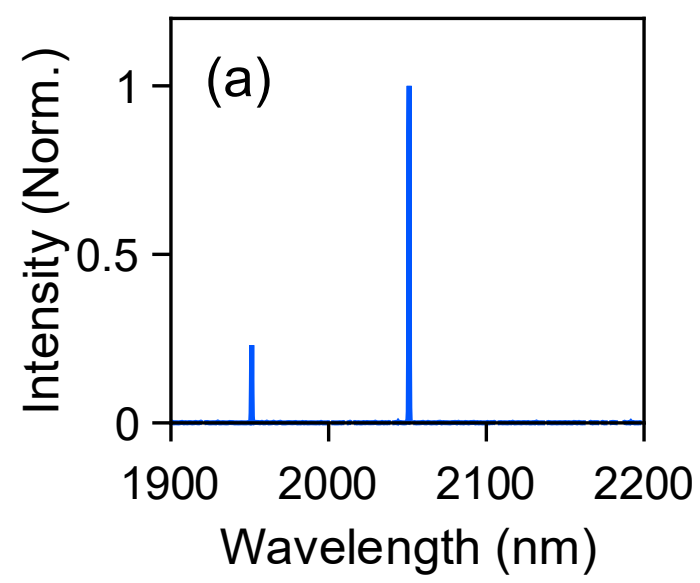

Fig. 4. Mode-locking results with DM1. (a) Laser spectrum before mode-locking initiation. (b) Incomplete mode-locking.

In the second step, DM1 was replaced by DM2. The CW lasing wavelength remained at ~2050 nm for all $T_{OC}$s and cavity working points. By slightly moving the HR mirror, stable mode-locking was successfully achieved at 6.3 W absorbed pump power. The average output power was 0.22 W. The corresponding laser spectrum is shown in Fig. 5(a). Fitting the measured spectrum with a sech² profile yields a central wavelength (λ) of 2084.3 nm and a full width at half-maximum (FWHM) bandwidth (Δλ) of 43.5 nm. The pulse duration was characterized using an autocorrelator (NX 50, A.P.E. GmbH). Assuming a sech² pulse shape, the fitted FWHM pulse duration (Δτ) was 107 fs. The resulting time–bandwidth product (TBP) was calculated to be 0.321, which is very close to the Fourier transform limit of 0.315. We further attempted to shorten the pulse duration by adjusting the GDD, but no obvious pulse shortening was achieved. This limitation is most likely attributed to the reduced available gain bandwidth from $Tm:Y_2O_3$ ceramic caused by DM2 [5], which restricts further pulse shortening and results in a relatively long pulse duration.

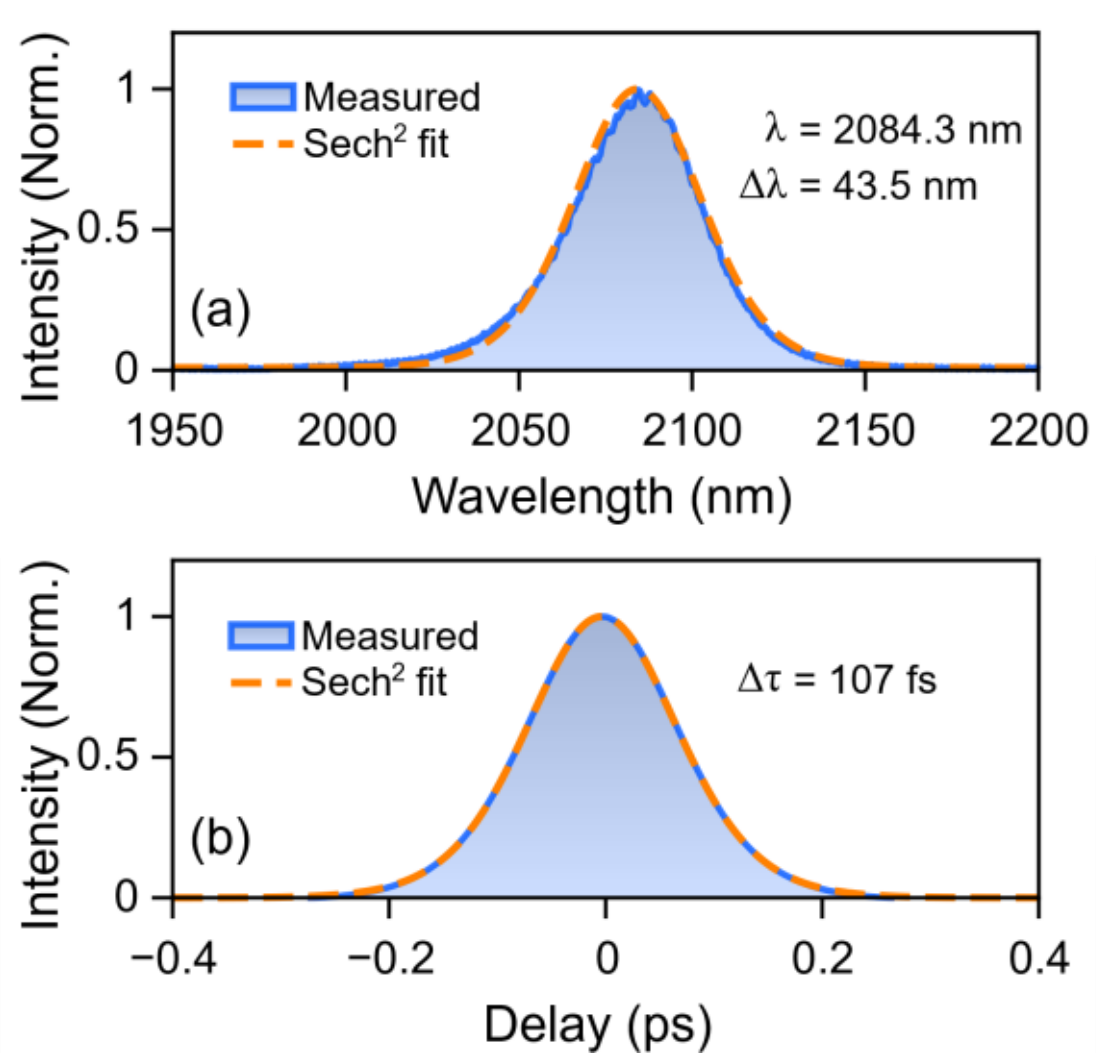

Fig. 5. Mode-locking results with DM2 and 1% $T_{OC}$. (a) Laser spectrum. (b) Non-collinear intensity autocorrelation trace.

Power scaling was achieved by increasing the $T_{OC}$. With 3% $T_{OC}$, stable mode-locking was achieved at a pump power of 8.2 W, delivering an average output power of 0.5 W. Figure 6(a) and 6(b) show the laser spectrum and autocorrelation trace, respectively. The central wavelength was 2081.3 nm, with an FWHM bandwidth of 41.6 nm. The pulse duration was 112 fs, corresponding to a TBP of 0.322. However, further increasing the $T_{OC}$ to obtain a higher output power failed to maintain stable mode-locking operation. This phenomenon can be attributed to the reduced instantaneous pulse intensity at a higher output coupling, which weakens the Kerr nonlinear effect and thereby hinders the initiation and stabilization of mode-locking.

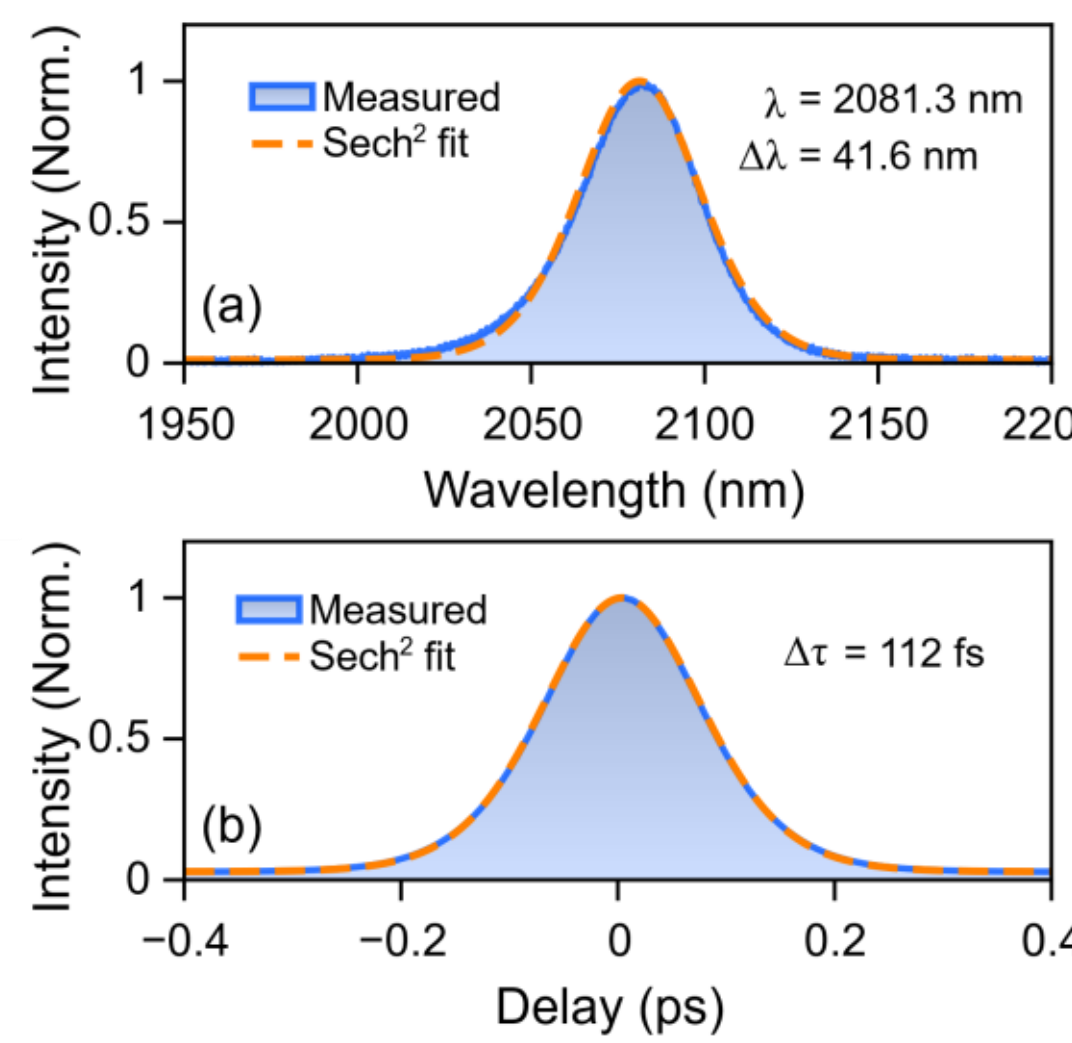

Fig. 6. Mode-locking results with DM2 and 3% $T_{OC}$. (a) Laser spectrum. (b) Non-collinear intensity autocorrelation trace.

The stability of the mode-locked pulses was characterized using a 12.5 GHz photodiode and a radio-frequency (RF) spectrum analyzer (Rigol DSA875). The RF spectrum of the mode-locking pulses at the highest out power is shown in Fig. 7. The beat signals over a span range of 1 GHz in Fig. 7(a) exhibit nearly identical amplitudes, confirming stable mode-locking. Figure 7(b) shows the fundamental RF beat note at a repetition rate of 90 MHz, with a spurious-free dynamic range of 60 dBc. The symmetric weak noise peaks located at 89.96 MHz and 90.04 MHz can be neglected since their amplitudes are lower than the amplitude-noise background [14].

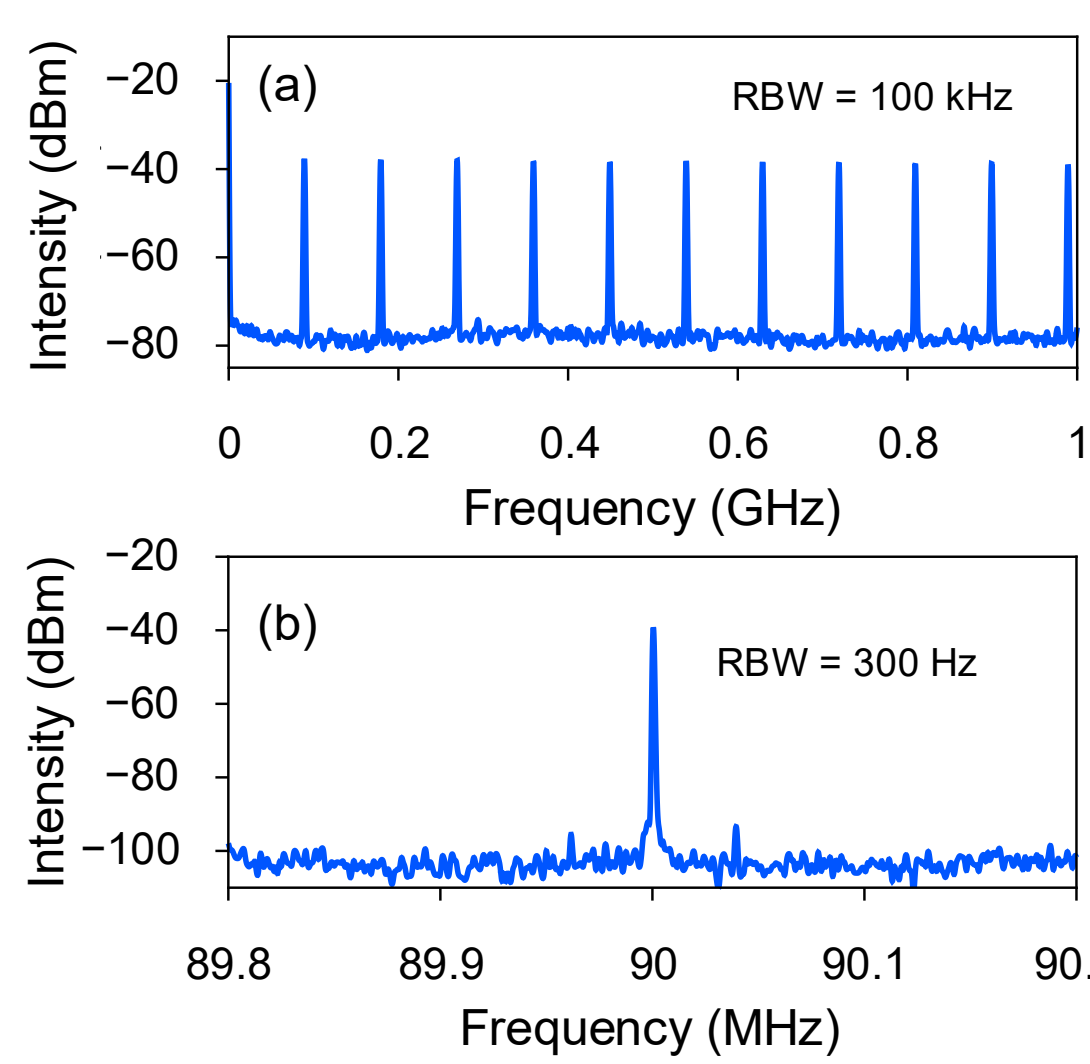

Fig. 7. RF spectra of the mode-locked $Tm:Y_2O_3$ laser. (a) Within a 1 GHz span; RBW, resolution bandwidth. (d) Fundamental beat signal within a 400 kHz span.

In summary, we have demonstrated pure KLM operation in a $Tm:Y_2O_3$ ceramic laser through precise spectral control using a custom-designed dichroic mirror with tailored reflection bands in the laser cavity. The shortest pulse duration of 107 fs was achieved at a central wavelength of 2084.3 nm with an average output power of 0.22 W. By further optimizing the output-coupling transmission, the average output power was increased to 0.5 W, with a slightly longer pulse duration of 112 fs and a central wavelength of 2081.3 nm. The shortest pulse duration was limited by the reflection bandwidth of the dichroic mirror. Further optimization of the mirror reflection spectrum–for instance, slightly blue-shifting to cover more gain band of $Tm:Y_2O_3$ while keeping high transmission at 1950 nm–together with adjustment of the pump diameter in the ceramic may achieve a shorter pulse duration while obtaining a higher output power.

The proposed spectral-control approach provides an effective solution for stabilizing KLM operation in gain media with multiple spectral peaks. Although SESAM-assisted KLM and the use of a much lower output coupling can also realize stable mode-locking [5, 8], our method is more conductive for power scaling and may serve as a versatile auxiliary technique for developing high-performance mode-locked lasers based on such gain materials.

**Funding.** This work was supported by the National Natural Science Foundation of China (62505331, 12388102), sponsored by Natural Science Foundation of Shanghai (25ZR1402532), and supported by the Strategic Priority Research Program of the Chinese Academy of Sciences (Grant No. XDB1940000).

**Disclosures**. The authors declare no conflicts of interest.

**Data availability.** Data underlying the results presented in this paper are not publicly available at this time but may be obtained from the authors upon reasonable request.